\documentclass[preprintnumbers,showpacs,amsmath,amssymb,prd,floatfix,twocolumn,superscriptaddress,nofootinbib]{revtex4}
\usepackage{graphicx}
\usepackage{epsfig}
\usepackage{bm}
\usepackage{amsfonts}

\newcommand\fverb{\setbox\pippobox=\hbox\bgroup\verb}
\newcommand\fverbdo{\egroup\medskip\noindent%
                        \fbox{\unhbox\pippobox}\ }
\newcommand\fverbit{\egroup\item[\fbox{\unhbox\pippobox}]}

\begin{document}

\title{Non-singular cosmology in a model of non-relativistic gravity}

\author{Yi-Fu Cai }
\email{caiyf@ihep.ac.cn} \affiliation{Institute of High Energy
Physics, Chinese Academy of Sciences, P.O. Box 918-4, Beijing
100049, P.R. China}

\author{Emmanuel N. Saridakis }
\email{msaridak@phys.uoa.gr} \affiliation{Department of Physics,
University of Athens, GR-15771 Athens, Greece}

\begin{abstract}
We present a model of non-relativistic gravitational theory which
is power-counting renormalizable in 3+1 dimensional spacetime.
When applied to cosmology, the relativity-violation terms lead to
a dark radiation component, which can give rise to a bounce if
dark radiation possesses negative energy density. Additionally, we
investigate a cyclic extension of the non-singular cosmology in
which the universe undergoes contractions and expansions
periodically. In both scenarios the background theory is well
defined at the quantum level.
\end{abstract}

 \maketitle

\section{Introduction}

Relativity is commonly assumed to be the foundation in
constructing models of particle physics and gravity. Yet, theories
based on relativity often suffer from some theoretical defects.
Namely, a quantum field theory of general relativity cannot be
well established since it is unable to be renormalized. Various
attempts to breaking relativity have been intensively discussed in
the literature. Recently, motivated by Ho\v{r}ava
\cite{Horava:2009uw, Horava:2008ih}, models of non-relativistic
quantum field theory were studied, not only theoretically
\cite{Calcagni:2009ar, Kiritsis:2009sh, Li:2009bg}, but also in
experimental detections \cite{Maccione:2009ju, Blas:2009my,
Gao:2009ht} (see also \cite{Mattingly:2005re}).

In the present work we are interested in constructing a model of
non-relativistic gravity and study its cosmological implications.
This model is power-counting renormalizable in 3+1 dimensions and
hence ultraviolet (UV) complete. Moreover, its action can recover
the exact Einstein-Hilbert form in the infrared (IR) limit, and so
general relativity and Lorentz symmetry in local frame do emerge
at low energy scales. A generic feature of this model is an
existence of dark radiation for which the energy density can be
either positive or negative. In the frame of a
Friedmann-Robertson-Walker (FRW) universe, dark radiation with
negative energy density can give rise to a bouncing solution
\cite{Brandenberger:2009yt}, since it breaks energy conditions if
accompanied by normal matter components \cite{Cai:2007qw}. This is
the so-called ``quintom scenario" \cite{Feng:2004ad, Li:2005fm,
Cai:2006dm, Setare:2008si} in which the equation-of-state of the
universe crosses the cosmological constant boundary. A remarkable
point of this model is that the violation of energy condition does
not bring the quantum instabilities \cite{Carroll:2003st,
Cline:2003gs} which often exist in usual quintom scenario. The
scenario of bouncing cosmology has been investigated in models
motivated by various approaches to quantum gravity
\cite{Veneziano:1991ek, Gasperini:1992em, Brustein:1997cv,
Bojowald:2001xe, Shtanov:2002mb, Apostolopoulos:2005ff}, and
analyzed using effective field techniques by introducing
energy-condition-violating matters \cite{Cai:2007qw,
Abramo:2007mp, Cai:2008qb, Cai:2008ed}, while the generation of
perturbation was studied in \cite{Mukhanov:1981xt, Hwang:2001zt,
Gasperini:2003pb, Peter:2002cn, Martin:2003sf, Piao:2003zm,
Allen:2004vz, Cai:2007zv} (and we refer to \cite{Novello:2008ra}
for a comprehensive review). One model of bouncing cosmology with
a matter-dominated contraction was found to be able to provide a
scale-invariant spectrum \cite{Starobinsky:1979ty, Wands:1998yp,
Finelli:2001sr, Peter:2006hx, Cai:2008qw} and sizable
non-Gaussianities \cite{Cai:2009fn, Cai:2009rd}, which may be
responsible for the current cosmological observations.

To extend, we find that the evolution of a universe in a model of
non-relativistic gravity might be free of any singularities. One
example of this scenario is an oscillating universe in which the
universe experiences a sequence of contractions and expansions
\cite{Xiong:2007cn, Xiong:2008ic, Saridakis:2009bv}. As shown in
\cite{Xiong:2008ic}, for a pivot process (bounce or turnaround) to
occur, one must require that the Hubble parameter vanishes while
its time derivative is non-vanishing at that point. Thus, it needs
to be satisfied that the equation-of-state $w$ of the universe is
much less than $-1$ at the bounce point while much larger than $1$
at turnaround point. Consequently, both the bounce and turnaround
break certain energy conditions. An oscillating universe realized
in a spatially flat universe has bee studied in
 \cite{Xiong:2008ic} by making use of a quintom matter with a
quantum instability. However, in the current work, we obtain a
cyclic scenario within a non-relativistic gravity model, which is
well defined at the quantum level but it merely requires that the
universe is not spatially flat.

This paper is organized as follows. In Section \ref{model} we
begin with the quantum theoretical puzzles of usual Einstein's
gravity, and we construct a non-relativistic gravitational model
which is power-counting renormalizable. In Section
\ref{application} we apply this model in the frame of an FRW
universe and then we obtain a bounce solution, in which we take a
canonical scalar field to mimic the normal matter component. In
addition, we find that the cosmological evolution in this
framework can be oscillatory and singularity-free, and the
corresponding form of the scalar-field potential can be
straightforwardly re-constructed. In Section \ref{Fluctuations} we
present a discussion on the cosmological fluctuations in the
model. Finally, in section \ref{concl} we summarize our results.

\section{A model of non-relativistic quantum gravity}
\label{model}

We start with a discussion on the main obstacle against the usual
approach to quantum gravity. In the context of quantum field
theory all successes are established on a solid construction of a
perturbatively renormalizable model, namely the SU(2) Standard
Model, in 3+1 dimensions. However, this procedure does not work in
the gravity sector, since the gravitational coupling constant $G$
possesses a negative dimension ($[G]=-2$ in mass units), which
introduces a UV incompletion to the theory.

The Einstein-Hilbert gravitational action of is given by
\begin{eqnarray}\label{SgEH}
S_g = \frac{1}{16\pi G} \int dt d^3x \sqrt{g}N \bigg(
K_{ij}K^{ij}-K^2 + R \bigg)~,
\end{eqnarray}
where
\begin{eqnarray}
K_{ij} = \frac{1}{2N} \left( {\dot{g_{ij}}} - \nabla_i N_j -
\nabla_j N_i \right) \, ,
\end{eqnarray}
is the extrinsic curvature and $R$ is the three-dimensional Ricci
scalar. The dynamical variables are the lapse and shift functions,
$N$ and $N_i$ respectively, and the spatial metric $g_{ij}$ (roman
letters indicate spatial indices), in terms of the ADM metric
\begin{eqnarray}
ds^2 = - N^2 dt^2 + g_{ij} (dx^i + N^i dt ) ( dx^j + N^j dt ) ,
\end{eqnarray} 
where indices are raised and lowered by $g_{ij}$.

Attempts on solving the UV incompletion have been intensively
studied in the literature. Motivated by a Lee-Wick model
\cite{Lee:1969fy, Lee:1970iw} which shows an improved UV behavior
\cite{Grinstein:2007mp}, gravity involving higher-derivative terms
may be applied to provide a UV completion. However, this approach
suffers from the existence of unbounded from below energy state,
and thus its quantization becomes unreliable. Another path to
quantum gravity is to build a non-local theory
\cite{Brandenberger:1988aj}, using infinite high-derivative terms
\cite{Biswas:2005qr, Biswas:2006bs, Barnaby:2008tc,
Calcagni:2009dg}, or string theory \cite{Freund:1987kt,
Freund:1987ck}, attempts which are still in proceeding.

Motivated by a recent work  \cite{Horava:2009uw}, one realizes
that a model of power-counting renormalizable gravity may be
achieved just by adding higher-order spatial derivative terms. One
important peculiarity in this type of models is that Lorentz
symmetry has to be given up but it may appear as an emergent one
at low energy scales. The original model, which is the so-called
Ho\v{r}ava gravity, since is required to satisfy a
`detailed-balance' condition referenced from condense matter
physics, it still suffers from problems such as the
over-constraining in UV region, being not compatible with current
observations even in the IR limit.

Concerning to above issues, the logic of effective field theory
suggests that a complete action of gravity could include all
possible terms consistent with the imposed symmetries, and the
dimensions of these terms ought to be bounded due to
renormalization. In the frame of 3+1 dimensional spacetime, a
renormalizable term may allow  6th-order spatial derivatives at
most, as pointed out by \cite{Horava:2009uw}. As a consequence,
one could add a modified action which involves all the permitted
terms:
\begin{eqnarray}\label{SgNR}
\Delta{S}_g&=&\frac{1}{16\pi G}\int dt d^3x \sqrt{g}N \bigg(
\alpha_1R_{ij}R^{ij}+\alpha_2R^2\nonumber\\
&&+\alpha_3\nabla_iR_{jk}\nabla^iR^{jk}+\alpha_4\nabla_iR_{jk}\nabla^jR^{ki}
\nonumber\\ &&+\alpha_5\nabla_i{R}\nabla^i{R} \bigg) ~.
\end{eqnarray}
Adding these terms into action (\ref{SgEH}) and requiring the
signs of the 6th-order spatial derivatives to be negative, we
obtain a non-relativistic gravity theory which is power-counting
renormalizable and the dispersion relation is bounded from below.
Note that in the following we are interested in the cosmological
applications of this gravitational background, that is in its IR
limit, and thus we do not present the full ``running'' formalism
but we restrict ourselves to the IR expressions. Furthermore,
without loss of generality, we will focus our investigation on the
quadratic terms, which preserves parity and Poincare symmetries,
although ones can straightforwardly examine a model with all the
coefficients included.

We can insert a matter component in the construction, attributed
to a canonical scalar field $\phi$  described by
\begin{eqnarray}\label{Sm}
S_m=\int dt d^3x \sqrt{g}N [
\frac{1}{2}\partial_\mu\phi\partial^\mu\phi-V(\phi)]~.
\end{eqnarray}
In addition, we focus on the cosmological frame with an FRW
metric,
\begin{eqnarray}
N=1~,~~g_{ij}=a^2(t)\gamma_{ij}~,~~N^i=0~,
\end{eqnarray}
with
\begin{eqnarray}
\gamma_{ij}dx^idx^j=\frac{dr^2}{1-k{r}^2}+r^2d\Omega_2^2~,
\end{eqnarray}
where $k=-1,0,1$ correspond to open, flat, and closed universe
respectively (with dimension $[k]=2$ in mass units). Both the
scalar field and the metric are assumed to be homogenous, with
their backgrounds being functions of cosmic time $t$. By varying
$N$ and $g_{ij}$, we obtain the Friedmann equations as follows
\begin{eqnarray}
\label{Fr1}
H^2&=&\frac{8\pi G}{3}\Big[\rho_m+\rho_{k}+\rho_{dr}\Big]~,\\
\label{Fr2} \dot{H}+\frac{3}{2}H^2&=&-4\pi
G\left[p_m-\frac{1}{3}\rho_{k}+\frac{1}{3}\rho_{dr}\right]~,
\end{eqnarray}
where $H\equiv\frac{\dot a}{a}$ is the Hubble parameter, and the
matter pressure and energy densities are expressed as
\begin{eqnarray}
\label{pm}
p_m&\equiv&\frac{\dot\phi^2}{2}-V(\phi)~,\\
\label{rhom}
\rho_m&\equiv&\frac{\dot\phi^2}{2}+V(\phi)~,\\
\rho_k&\equiv&-\frac{3k}{8\pi Ga^2}~,\label{rhok}\\
\rho_{dr}&\equiv&-\frac{3(\alpha_1+3\alpha_2)k^2}{4\pi Ga^4}~,
\label{rhodr}
\end{eqnarray}
respectively. In addition, the energy density for matter component
satisfies the continuity equation $\dot{\rho}_m+3H(\rho_m+p_m)=0$
following from the action (\ref{Sm}), which leads to the equation
of motion for the scalar field
\begin{eqnarray}\label{eom}
\ddot\phi+3H\dot\phi+V_{,\phi}=0~,
\end{eqnarray}
where the subscript `$_{,\phi}$' denotes the derivative with
respect to $\phi$.

In summary, in Friedmann equation (\ref{Fr1}) we have obtained the
energy density for usual matter component (with equation-of-state
parameter $-1\leq w_m\leq1$, where $w_m\equiv p_m/\rho_m$) and the
usual spatial curvature term (of which the equation-of-state
parameter is $w_{k}=-1/3$ as can be seen from (\ref{Fr2})).
Besides, we have derived a ``dark radiation term'', which evolves
proportionally to $a^{-4}$ and thus its equation-of-state
parameter is $w_{dr}=1/3$ (as can be seen from (\ref{Fr2})). This
last term reflects a novel feature of a model of non-relativistic
gravity, leading to smooth cosmological evolutions  with the
initial singularity replaced by a big bounce, if $w_m<w_{dr}$ and
$\alpha_1+3\alpha_2>0$ and the universe is not exactly flat.

\section{Cosmological Evolutions}
\label{application}

In this section we will study in detail the background evolution
in the context of non-relativistic gravity. In particular, we
study a non-flat universe with $k=\pm1$ and we define a critical
energy density as
$\rho_{cr}\equiv\frac{3(\alpha_1+3\alpha_2)k^2}{4\pi G}$ which
needs to be larger than zero. Furthermore, we normalize $a=1$ at
the bounce point.

\subsection{A bounce solution}

Let us take a first look at how it is possible to obtain a
bouncing cosmology in this model. We assume that the universe
starts in a contracting phase and that the contribution of the
normal matter dominates over that of the dark radiation. This will
typically be the case at low energy densities and curvatures. As
the universe contracts and the energy density increases, the
relative importance of dark radiation compared to the normal
matter will grow. From (\ref{Fr1}) it follows that there will be a
time when $H = 0$. This is a necessary condition for the
realization of the bounce. By making use of the continuity
equations it follows that at the bounce point $\dot H
> 0$. Hence, we can acquire a transition from a contracting to
an expanding phase, which is a cosmological bounce.

Now, as a specific example, for the scalar field responsible for
matter we consider a mass potential
$V(\phi)=\frac{1}{2}m^2\phi^2$. We begin the evolution during the
contracting phase with a sufficiently large scale factor, so that
we expect the contribution of the dark radiation to the total
energy density to be small. Besides we require the contribution of
the spatial curvature term to be negligible too (we leave the
discussion on this component for the next section). For these
initial conditions the scalar field will be oscillating around its
vacuum, and the equation of state will hence be that of a matter
dominated universe with its Hubble parameter being
\begin{eqnarray}
\langle H \rangle \simeq \frac{2}{3t}~
\end{eqnarray}
on average. In fact, as follows from the Klein-Gordon equation
(\ref{eom}), which in this case reduces to
\begin{eqnarray}
\ddot\phi+3H\dot\phi+m^2\phi=0~,
\end{eqnarray}
 one well-known approximated solution is
\begin{eqnarray}
\phi(t)\simeq \frac{1}{\sqrt{3\pi G}}\frac{\sin {mt}}{mt}~.
\end{eqnarray}

When $t$ is smaller than $m^{-1}$, the scalar field $\phi$ is
oscillating and the universe behaves as a matter-dominated one.
Once   $t\sim m^{-1}$, the amplitude of $\phi$ reaches the Planck
scale and the scalar enters a ``slow-climb" phase. During this
period, the energy density of the scalar varies very slowly but
the scale factor decreases rapidly. Therefore, the dark radiation
will become similar to the scalar field density very soon. Once
there is $\rho_m=\rho_{cr}$, we will obtain a big bounce. Since in
this case  the universe has experienced a matter-dominated
contraction, this scenario is the so-called ``matter-bounce".

\subsection{Cyclic scenario} \label{Cyclic}

Having investigated the realization of bouncing cosmology in the
present non-relativistic gravitational framework, we now study a
sub-class of cosmological evolution without any singularities,
that is a realistic and physical cyclic scenario.

A cyclic scenario could be straightforwardly obtained if we
consider a negative dark radiation term and a negative curvature
one. According to (\ref{rhok}) and (\ref{rhodr}), this is achieved
if $\alpha_1+3\alpha_2>0$ and if $k=1$, that is for a closed
universe. For simplicity we consider the matter component of the
universe to be dust, that is to possess $w_m=0$. During the
expansion, the energy densities of all components are decreasing.
However, the energy density of the curvature term decreases much
slower relatively to others. Thus, its contribution will
counterbalance that of dark matter, triggering a turnaround with
$H=0$ and $\dot H<0$, after which the universe enters in the
contracting phase. As described in the previous subsection, after
a contraction to sufficiently small scale factors the dark
radiation term will lead the universe to experience a bounce.
Therefore, the universe in such a model indeed presents a cyclic
behavior, with a bounce and a turnaround at each cycle.

A more careful analysis reveals that the negative curvature term,
that is a closed universe with $k=1$, is not a necessary
condition. Indeed, its role on the competition of the positive
dark matter density can be equivalently fulfilled by a small
negative cosmological constant added to the potential $V(\phi)$.
In other words, a slightly negative $V(\phi)$ can trigger the
turnaround at large scale factors, even if the universe is open,
i.e with $k=-1$.

The above general examples reveal the realization of a
singularity-free cyclic universe in a framework of
non-relativistic gravity. This behavior can be obtained for both
closed and open universe, and the only restriction is the
requirement of a matter content with equation-of-state parameter
between $-\frac{1}{3}$ and $\frac{1}{3}$. Thus, having extracted
its general features, in the following we will explicitly
construct the class of models that allow for cyclicity, and in
particular we wish to appropriately re-construct the corresponding
scalar potential.

Let us first start from the desired result, that is to impose a
known scale factor $a(t)$ possessing an oscillatory behavior. In
this case both $H(t)$ and $\dot{H}(t)$ are straightforwardly
known. Therefore, we can use the the Friedmann equations
(\ref{Fr1}),(\ref{Fr2}) together with (\ref{pm}),(\ref{rhom}), in
order to extract the relations for $\phi(t)$ (through
$\dot{\phi}(t)$) and $V(t)$, acquiring:
\begin{eqnarray}
 \label{phidot3}
 \phi(t) &=& \pm \int^tdt' \sqrt{-\frac{\dot H(t')}{4\pi G}-\frac{2}{3}\rho_{k}(t')-\frac{4}{3}\rho_{dr}(t')},\ \ \ \\
 \label{Vt3}
 V(t) &=& \frac{\dot H(t)}{8\pi G}+\frac{3H^2(t)}{8\pi G}-\frac{2}{3}\rho_{k}(t)-\frac{1}{3}\rho_{dr}(t)~.
\end{eqnarray}
Note that the $a(t)$-form or the parameter-choices must lead to a
positive $\dot{\phi}^2(t)$. Finally, eliminating time between
these two expressions we extract the explicit form of the
potential $V(\phi)$. Thus, performing the procedure inversely we
conclude that this particular $V(\phi)$ generates the desired
$a(t)$-form.

We now proceed to a specific, simple, but quite general example.
We assume a cyclic universe with an oscillatory scale factor of
the form
\begin{equation} \label{at}
a(t)=A\sin(\omega t)+a_c,
\end{equation}
where we have shifted $t$ in order to eliminate a possible
additional parameter standing for the phase. Furthermore, the
non-zero constant $a_c$ is inserted in order to eliminate any
possible singularities from the model. In such a scenario $t$
varies between $-\infty$ and $+\infty$, and $t=0$ is just a
specific moment without any particular physical meaning. Finally,
note that the bounce occurs at $a_{B}(t)=a_c-A$, which can be set
to $1$. Straightforwardly we find:
\begin{eqnarray} \label{Ht}
&&H(t)=\frac{A\omega\cos(\omega t)}{A\sin(\omega t)+a_c}\\
\label{Hpt} && \dot{H}(t)=-\frac{A\omega^2\left[A+a_c\sin(\omega
t)\right]}{\left[A\sin(\omega t)+a_c\right]^2},
\end{eqnarray}
and thus substitution into (\ref{rhok}),(\ref{rhodr}),
(\ref{phidot3}) and (\ref{Vt3}) gives the corresponding
expressions for $\phi(t)$ and $V(t)$.

\begin{figure}[ht]
\begin{center}
\mbox{\epsfig{figure=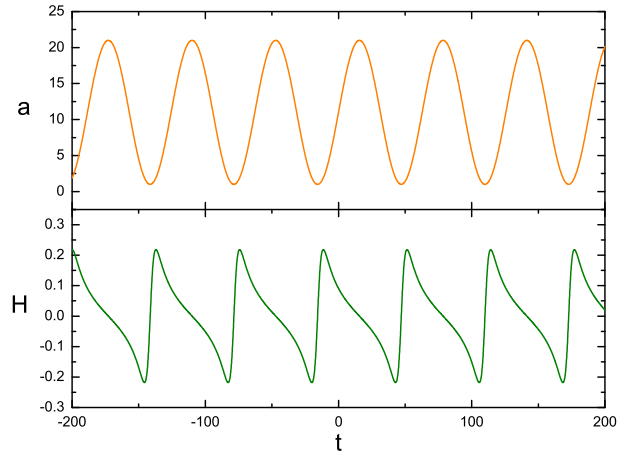,width=9.3cm,angle=0}} \caption{{\it
The evolution of the scale factor $a(t)$ and of the Hubble
parameter $H(t)$ of the ansatz (\ref{at}), with $A=10$,
$\omega=0.1$ and $a_c=11$. All quantities are measured in units
where $8\pi G=1$.
 }} \label{fig1}
\end{center}
\end{figure}

\begin{figure}[ht]
\begin{center}
\mbox{\epsfig{figure=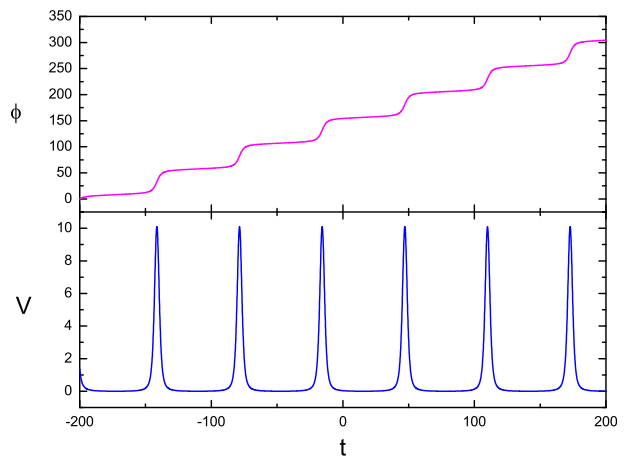,width=9.3cm,angle=0}} \caption{{\it
$\phi(t)$ and $V(t)$ for the cosmological evolution of
Fig.~\ref{fig1}, in the case of a closed universe ($k=1$) with
$\alpha_1=1$ and $\alpha_2=1$. All quantities are measured in
units where $8\pi G=1$.}} \label{fig2}
\end{center}
\end{figure}

\begin{figure}[ht]
\begin{center}
\mbox{\epsfig{figure=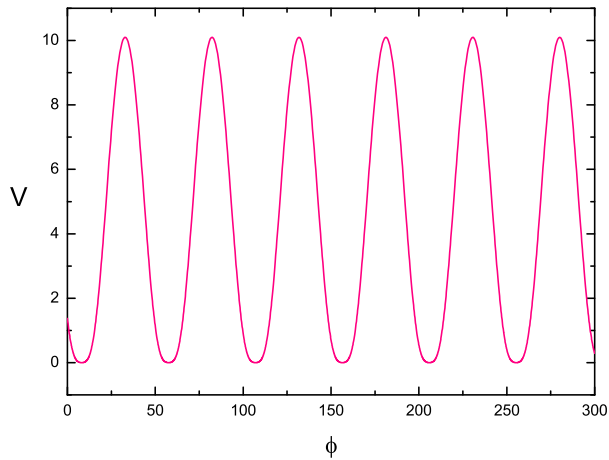,width=9.3cm,angle=0}} \caption{{\it
$V(\phi)$ for the cosmological evolution of Figs.~\ref{fig1} and
\ref{fig2}. All quantities are measured in units where $8\pi
G=1$.}} \label{fig3}
\end{center}
\end{figure}

In order to provide a more transparent picture of the obtained
cosmological behavior, in Fig.~\ref{fig1} we present the evolution
of the scale factor (\ref{at}) and of the Hubble parameter
(\ref{Ht}) with $A=10$, $\omega=0.1$ and $a_c=11$, where all
quantities are measured in units with $8\pi G=1$. This choice is
consistent with our setting $a=1$ at the bounce.
In Fig.~\ref{fig2} we depict the corresponding behavior of
$\phi(t)$ and $V(t)$ for the scale factor of Fig.~\ref{fig1}, in
the case of a closed universe ($k=1$), and for model parameters
$\alpha_1=1$ and $\alpha_2=1$ (in units where $8\pi G=1$).
Finally, eliminating time between $\phi(t)$ and $V(t)$ allows us
to re-construct the corresponding relation for $V(\phi)$, shown in
Fig.~\ref{fig3}.

From these figures we observe that an oscillating and
singularity-free scale factor, can be generated by an oscillatory
form of the scalar potential $V(\phi)$ (although of not a simple
function as that of $a(t)$, as can be seen by the slightly
different form of $V(\phi)$ in its minima and its maxima). This
$V(\phi)$-form was more or less theoretically expected, since a
non-oscillatory $V(\phi)$ would be physically impossible to
generate an infinitely oscillating scale factor and a universe
with a form of time-symmetry. Finally, we stress that although we
have presented the above specific simple example, we can
straightforwardly perform the described procedure imposing an
arbitrary oscillating ansatz for the scale factor.

The aforementioned bottom to top approach was enlightening about
the form of the scalar potential that leads to a cyclic
cosmological behavior. Therefore, one can perform the above
procedure the other way around, starting from a specific
oscillatory $V(\phi)$ and resulting to an oscillatory $a(t)$. In
particular, (\ref{phidot3}) is written in a compact form as
$\dot{\phi}^2(t)=Q_1(a,\dot{a},\ddot{a})$ and similarly
(\ref{Vt3}) as $ V(t)=Q_2(a,\dot{a},\ddot{a})$. Thus, we can
invert  the known form of $V(\phi)\equiv V(\phi(t))$ obtaining
$\phi(t)=V^{\{-1\}}\left(Q_2(a,\dot{a},\ddot{a})\right)$.
Therefore,
$\dot{\phi}^2(t)=\left\{\frac{d}{dt}\left[V^{\{-1\}}\left(Q_2(a,\dot{a},\ddot{a})\right)\right]\right\}^2$.
In conclusion, the scale factor arises as a solution of the
differential equation
\begin{equation} \label{Diffat}
Q_1(a,\dot{a},\ddot{a})=\left\{\frac{d}{dt}\left[V^{\{-1\}}\left(Q_2(a,\dot{a},\ddot{a})\right)\right]\right\}^2.
\end{equation}
As a specific example we consider the simple case
\begin{equation} \label{Vtspec}
V(\phi)=V_0\sin(\omega_V\, \phi)+V_c.
\end{equation}
In this case
$\phi=\frac{1}{\omega_V}\sin^{-1}\left(\frac{V(\phi(t))-V_c}{V_0}\right)$,
where $V(\phi(t))\equiv V(t)=Q_2(a,\dot{a},\ddot{a})$  with
$Q_2(a,\dot{a},\ddot{a})$ the right hand side of expression
(\ref{Vt3}). Therefore, differentiation leads to:
\begin{equation}
\dot{\phi}(t)=\frac{1}{V_0\,\omega_V}\frac{1}{\sqrt{1-\left[\frac{Q_2(a,\dot{a},\ddot{a})-V_c}{V_0}\right]^2}}
\,\frac{d}{dt}\left[Q_2(a,\dot{a},\ddot{a})\right]
\end{equation}
and thus we obtain
\begin{equation} \label{Diffeqa}
Q_1(a,\dot{a},\ddot{a})=\left\{\frac{1}{V_0\,\omega_V}\frac{1}{\sqrt{1-\left[\frac{Q_2(a,\dot{a},\ddot{a})-V_c}{V_0}\right]^2}}
\,\frac{d}{dt}\left[Q_2(a,\dot{a},\ddot{a})\right]\right\}^2,
\end{equation}
where as we have mentioned, $Q_1(a,\dot{a},\ddot{a})$ is the right
hand side of expression (\ref{phidot3}).

\begin{figure}[ht]
\begin{center}
\mbox{\epsfig{figure=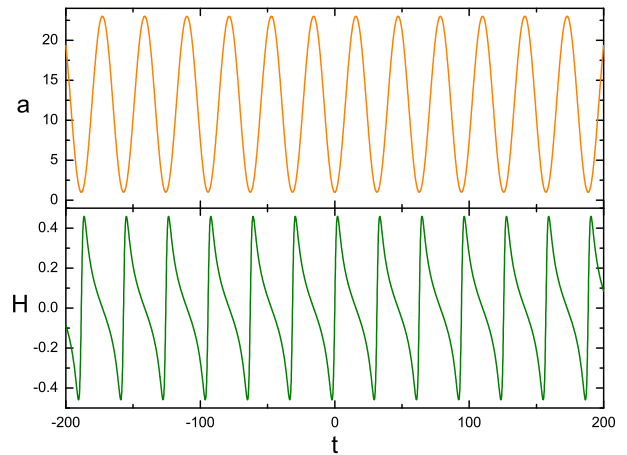,width=9.3cm,angle=0}} \caption{{\it
The evolution of the scale factor $a(t)$ and of the Hubble
parameter $H(t)$, for a scalar potential of the ansatz
(\ref{Vtspec}) with $V_0=5.25$, $\omega_V=0.25$ and $V_c=5.25$,
with $k=1$, $\alpha_1=1$ and $\alpha_2=1$.  All quantities are
measured in units where $8\pi G=1$.}} \label{fig4}
\end{center}
\end{figure}

Differential equation (\ref{Diffeqa}) cannot be handled
analytically, but it can be easily solved numerically. In
Fig.~\ref{fig4} we depict the corresponding solution for $a(t)$
(and thus for $H(t)$) under the ansatz (\ref{Vtspec}) with
$V_0=5.25$, $\omega_V=0.25$ and $V_c=5.25$, with $k=1$,
$\alpha_1=1$ and $\alpha_2=1$ (in units where $8\pi G=1$). The
potential parameters have been chosen in order to acquire a cyclic
universe with $a(t)\approx1$ at the bounce.

\begin{figure}[ht]
\begin{center}
\mbox{\epsfig{figure=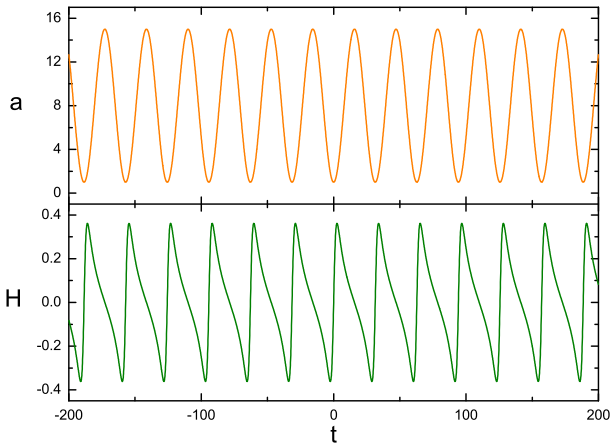,width=9.3cm,angle=0}} \caption{{\it
The evolution of the scale factor $a(t)$ and of the Hubble
parameter $H(t)$, for a scalar potential of the ansatz
(\ref{Vtspec}) with $V_0=3.15$, $\omega_V=0.25$ and $V_c=3.13$,
with $k=-1$, $\alpha_1=1$ and $\alpha_2=1$.  All quantities are
measured in units where $8\pi G=1$.}} \label{fig5}
\end{center}
\end{figure}

Let us now present the corresponding simple cyclic example in the
case of an open universe ($k=-1$). In Fig.~\ref{fig5} we depict
the evolutions of the scale factor $a(t)$ and thus of the Hubble
parameter $H(t)$, under the ansatz (\ref{Vtspec}) with $V_0=3.15$,
$\omega_V=0.25$ and $V_c=3.13$, with $k=-1$, $\alpha_1=1$ and
$\alpha_2=1$ (in units where $8\pi G=1$).
As described above, in the case of an open universe one needs the
scalar potential to be negative for field values corresponding to
large scale factors, in order for the turnaround to be triggered.
However, since at that regime the curvature term is very small,
even very small negative potential values can fulfill this
condition, as can be seen by the specific example of
Fig.~\ref{fig5}, where $V_0=3.15$ and $V_c=3.13$ leading the
minimal value of the potential to be $-0.02$.

Note that the specific examples of the above figures are just
simple representatives of cyclic behavior in our gravitational and
cosmological construction, and they correspond only to a sub-class
of the whole set of cyclic models. Obviously, one can
straightforwardly generalize the aforementioned procedure in any
periodic model. For instance, imposing any periodic function for
$a(t)$ one results in the corresponding periodic scalar potential
$V(\phi)$. Similarly, imposing any periodic potential $V(\phi)$
one can solve the differential equation (\ref{Diffat}) and extract
the resulting periodic $a(t)$.

We close this section by mentioning that, although the bounce
solutions arise owing to the presence of a dark radiation
component with negative energy density, they can also be obtained
if ordinary radiation with positive energy density is present.
When ordinary radiation is involved, it has to be generated from
the reheating process of a primordial field namely inflaton or
$\phi$ appeared in our model. Therefore its domination only takes
place after reheating of which the energy scale is much lower than
the bounce scale. Moreover, in the late time evolution normal
radiation would be erased during matter dominated period, and
hence will not affect the bounce solution in next cycle. We will
address on the details of this issue in future studies.

\section{Fluctuations through the bounce}
\label{Fluctuations}

A model of non-relativistic gravity is usually able to recover
Einstein's general relativity as an emergent theory at low energy
scales. Therefore, the cosmological fluctuations generated in this
model should be consistent with those obtained in standard
perturbation theory in the IR limit \cite{Brandenberger:2009yt}.
This result has been intensively discussed in the literature (see
e.g. \cite{Cai:2007zv}). In particular, the perturbation spectrum
presents a scale-invariant profile if the universe has undergone a
matter-dominated contracting phase \cite{Cai:2008ed, Cai:2008qw,
Cai:2009rd}. However, the non-relativistic corrections in the
action (\ref{SgNR}) could lead to a modification of the dispersion
relations of perturbations. This issue has been addressed in
\cite{Cai:2009hc} \footnote{we refer to Refs.
\cite{Mukohyama:2009gg, Piao:2009ax} and references therein for
the perturbations of a pure expanding universe in
Ho\v{r}ava-Lifshitz cosmology.}, which shows that the spectrum in
the UV regime may have a red tilt in a bouncing universe.
Moreover, the perturbation modes cannot even enter the UV regime
in the scenario of matter-bounce. So the analysis of the
cosmological perturbations in the IR regime is quite reliable.

Things become complicated but more interesting in a cyclic
scenario. Usually, a particular perturbation mode in the
contracting phase is dominated by its growing tendency, but in the
expanding stage it becomes nearly constant on super-Hubble scales.
Therefore, the metric perturbation is amplified on super-Hubble
scales cycle by cycle \cite{Piao:2009ku}, and also the slope of
its spectral index is varying \cite{Brandenberger:2009ic}.
However, it is known that the contribution of fluctuations has to
be much less than the background energy. This prohibits the metric
perturbations to enter the next cycle if $\delta\rho/\rho\sim
O(1)$, unless the universe can be separated into many parts
independent of one another, each of which corresponding to a new
universe and evolving up to next cycle, then separate again and so
on. In this case, the model of cyclic universe may be viewed as a
realization of the multiverse scenario \cite{Erickson:2006wc,
Lehners:2008qe, Piao:2009ku}.

\section{Conclusions}
\label{concl}

In this work, we have studied the possibility of constructing a
model of non-relativistic quantum gravity in 3+1 dimensional
spacetime. The novel features of the gravitational sector are
reflected in new terms that are present in the IR, that is in the
cosmologically interesting regime. Our results show that this
model can give rise to a non-singular cosmology, for which the
initial singularity is replaced by a big bounce if we require that
the dark radiation term is negative. Specifically, we have
considered an example in which the normal matter component is a
free scalar field. In this case we have obtained a bouncing
universe with a matter-dominated contracting phase, and so it may
be responsible for the formation of a scale-invariant primordial
spectrum. To extend, we have also investigated the realization of
a cyclic scenario, by re-constructing the potential of the scalar
field which leads to a universe with an oscillatory scale factor.

Recently, the scenario of oscillating universe (originally
proposed by \cite{tolman} and awaked later as ekpyrotic/cyclic by
\cite{Steinhardt:2002ih, Khoury:2001bz, Tsujikawa:2002qc}), in
which the universe experiences a sequence of contractions and
expansions, has been widely studied in the literature, for
instance in the context of loop quantum gravity
\cite{Bojowald:2004kt, Lidsey:2004ef, Singh:2006im, Xiong:2007cn},
including matter components violating energy conditions
\cite{Brown:2004cs, Feng:2004ff, Dabrowski:2004hx, Baum:2006nz,
Clifton:2007tn, Freese:2008pu}, in the frame of string cosmology
\cite{Steinhardt:2001st, Steinhardt:2002ih} and within the
brane-world \cite{Kanekar:2001qd, Shtanov:2002mb,Saridakis:2007cf}
(see Refs. \cite{Piao:2004me, Zhang:2007yu, Biswas:2008kj,
Biswas:2009fv, Liu:2009nv, Nozari:2009jy, Zhang:1900zz,
Zhang:2009xp, Koshelev:2009ty} for recent developments on
oscillating universes). The main peculiarity of the current work
is that the oscillation depends on the dark radiation term, which
is present only for  not spatially flat geometry. However, the
main advantage is that the background theory is well defined at
the quantum level.

In conclusion, we see that the present model of power-counting
renormalizable, non-relativistic gravitational theory in 3+1
dimensional spacetime, can naturally lead to a bounce and to
cyclicity as particular sub-classes of its possible induced
cosmological behaviors. The fact that the theory is UV complete
and that instabilities do not arise at the quantum level, makes
future investigations on its cosmological implications quite
interesting.

As an end, we would like to comment on a principal difference
between our model and the intensely studied Ho\v{r}ava-Lifshitz
cosmology\cite{Calcagni:2009ar, Kiritsis:2009sh}. In a model of
Ho\v{r}ava-Lifshitz cosmology (with or without detailed balance
conditions), it is claimed by \cite{Charmousis:2009tc} that, there
is one extra scalar modes in its perturbation theory which becomes
strongly coupled as the parameters approach a desired IR fixed
point. To understand this point, we would like to recall that such
a strong coupling problem only exists in a system of which the
original scalar modes have obtained effective mass terms and so
lead to the generation of an extra degree of freedom, as what we
have understood in quantum field theory very well. As a
consequence, modifications on general relativity often suffers
from this problem, namely, in the theory of Pauli-Fierz massive
gravity \cite{Fierz:1939ix} the longitudinal scalar becomes
strongly coupled when the mass approaches zero
\cite{ArkaniHamed:2002sp}, leading to the famous vDVZ
discontinuity \cite{vanDam:1970vg, Zakharov:1970cc}.
Ho\v{r}ava-Lifshitz cosmology also suffers from the strong
coupling problem, since the theory manifestly contains
parity-violating terms\cite{Takahashi:2009wc, Bogdanos:2009uj}
which bring effective masses for gravitons. However, this does not
happen in our model since the spatial curvature terms and its
spatial derivatives introduced phenomenologically only contribute
to the dispersion relations of two polarizations of gravitons
without bringing effective mass terms. In this case, there is no
extra degree of freedom in our model and the longitudinal part of
metric perturbations can be fixed by Hamiltonian constraint when
combined with matter component.

\begin{acknowledgments}
We would like to thank Xinmin Zhang for discussions. The research
of Y.F.C. is supported in part by the National Science Foundation
of China under Grants No. 10533010 and 10675136, and by the
Chinese Academy of Sciences under Grant No. KJCX3-SYW-N2.
\end{acknowledgments}

\end{document}